\documentclass[apj]{emulateapj}

\shorttitle{Radio emission of Sw 1644+57}
\shortauthors{Barniol Duran \& Piran}

\usepackage{epsf}
\usepackage{epsfig}
\usepackage{amsmath}
\usepackage{color} 
\usepackage{ulem}

\newcommand{\gae}{\lower 2pt \hbox{$\, \buildrel {\scriptstyle >}\over {\scriptstyle
\sim}\,$}}
\newcommand{\lae}{\lower 2pt \hbox{$\, \buildrel {\scriptstyle <}\over {\scriptstyle
\sim}\,$}}

\begin{document}

\title{On the origin of the radio emission of Sw 1644+57}
\author{Rodolfo Barniol Duran\altaffilmark{1a} \& Tsvi Piran\altaffilmark{1b}}
\altaffiltext{1}{Racah Institute for Physics, The Hebrew University, Jerusalem, 91904, Israel}
\email{(a) rbarniol@phys.huji.ac.il; (b) tsvi.piran@mail.huji.ac.il}

\begin{abstract}
We apply relativistic equipartition synchrotron arguments to the puzzling 
radio emission of the tidal disruption event candidate Sw 1644+57. We find 
that, regardless of the details of the equipartition scenario considered, 
the energy required to produce the observed radio (i.e., energy in the
magnetic field and radio emitting electrons) must increase by a factor of
$\sim 20$ during the first 200 days. It then saturates. This energy increase cannot
be alleviated by a varying geometry of the system. The radio data can be 
explained by: (i) An afterglow like emission of the X-ray emitting narrow 
relativistic jet. The additional energy can 
arise here from a slower moving material ejected in the first few days that 
gradually catches up with the slowing down blast wave (Berger et al. 2012). 
However, this requires at least $\sim 4 \times 10^{53}$ erg in the slower moving 
outflow. This is much more than the energy of the fast moving outflow that 
produced the early X-rays and it severely constrains the overall energy 
budget. (ii) Alternatively, the radio may arise from a mildly relativistic 
and quasi-spherical outflow. Here, the energy available for radio emission increases with time reaching 
at least $\sim 10^{51}$ erg after 200 days. This scenario requires, however, 
a second separate X-ray emitting collimated relativistic component. Given these 
results, it is worthwhile to consider alternative models in which energy of
the magnetic field and/or of the radio emitting electrons increases with time 
without having a continuous energy supply to the blast wave. This can happen, 
for example, if the energy is injected initially mostly in one form (Poynting 
flux or baryonic) and it is gradually converted to the other form, leading to 
a strong time-varying deviation from equipartition. Another intriguing
possibility is that a gradually decreasing Inverse Compton cooling modifies 
the synchrotron emission and leads to an increase of the available energy in
the radio emitting electrons (Kumar et al. 2013).
\end{abstract}

\section{Introduction}

Swift J164449.3+573451 (hereafter, Sw 1644+57), a peculiar high energy transient,
was detected at the end of March 2011. It has been interpreted as the tidal
disruption of a main-sequence star by a supermassive black hole (e.g., Bloom et al. 2011,
Burrows et al. 2011, Levan et al. 2011, Zauderer et al. 2011, 2013, Berger et
al. 2012).  Krolik \& Piran (2011) found that a white dwarf tidal disruption
provides several advantages in explaining the early time X-ray data over the 
disruption of a main-sequence star. Alternative interpretations include a very long
Gamma-Ray Burst (Quataert \& Kasen 2012) and a quark-nova (Ouyed, Staff \& Jaikumar 2011). 

Sw 1644+57 was detected by the {\it Swift} Burst Alert Telescope and 
X-Ray Telescope as a strongly flaring radio transient that $\sim 10$ days after
the trigger began to decrease roughly as $\propto t^{-5/3}$ 
(Bloom et al. 2011, Burrows et al. 2011).  The X-ray emission was variable and 
continued to decrease monotonically until $\sim 600$ days, when it abruptly
underwent a sharp decline in the X-ray flux as
detected by {\it Chandra} (Zauderer et al. 2012). 
Radio observations began $\sim 5$ days after the
trigger (Zauderer et al. 2011, Wiersema et al. 2012) and, since then, 
a long-term program to monitor this event with several radio
facilities is currently underway (Berger et al. 2012, 
Zauderer et al. 2013).  The radio emission shows a
self-absorbed synchrotron spectrum and smooth light curves. 
Sw 1644+57 was also monitored by several ground telescopes in the optical and
near infrared (and also with HST), which allowed a determination of the redshift
of $z=0.354$ (Levan et al. 2011).

Zauderer et al. (2011; hereafter, Z11) used equipartition arguments to
interpret the observed radio emission from 5--22 days after the initial
detection.  They suggest that this radio emission arises from a mildly
relativistic outflow with a constant Lorentz Factor (LF) of 
$\Gamma \approx 1.2$.  Z11 found the following properties using the radio data: 
1. The radio emission was not produced by the same relativistic electrons
that produced the X-ray emission, 2. The external density medium profile 
is approximately $\propto R^{-2}$, 3. The total energy increased by a factor of $\sim 2$
over the time span of these observations.  Metzger, Giannios \& Mimica (2012;
hereafter, MGM12) suggested an ``afterglow'' model, in which they
interpret the radio observations in Z11 as synchrotron radiation produced by
the shock interaction between the jet that has produced the X-rays 
and the external medium.  They also
find a density medium profile as $\propto R^{-2}$, however, they find
that the outflow must be narrowly collimated with opening angle smaller than $1/\Gamma$,
which was different compared with the geometry used by Z11.
Later, continuing with their observational campaign, Berger et al. (2012;
hereafter, B12) presented radio data on a longer time
scale of 5--216 days and used the MGM12 model to fit their data.  
Significant deviations from the predictions of the MGM12
model were needed to explain the radio observations. In particular, B12, who use the same
theoretical model as MGM12, namely, a narrowly collimated jet, 
find that the total energy increased by about an
order of magnitude from 5 to $\sim 200$ days, and the overall density profile exhibits a
significant flattening at a larger radius (see, also, Cao \& Wang 2012). Zauderer et al. (2013; hereafter,
Z13) continued the observational campaign until $\sim 600$ days and found no
deviations from the conclusions found in B12 concerning the radio data.

The increase of energy is puzzling as there is no indication for an additional
energy injection from the X-ray emission, which was mostly
emitted within the first $\sim 3$ days (Burrows et al. 2011).  B12 suggested that
this energy increase results from energy coming from a slower moving matter that 
catches up with the shock at a
later stage (alternatively, see, De Colle et al. 2012, Liu, Pe'er \& Loeb 2012, 
Tchekhovskoy et al. 2013).  The density profile is also
somewhat surprising and B12 interpret it as a possible indication of Bondi
accretion.  In order to explore the robustness of these conclusions
and to assess if these features arise from the specific assumptions of the MGM12 afterglow
model, we apply here our recently developed generalized relativistic
equipartition arguments (Barniol Duran, Nakar, Piran 2013, hereafter ``Paper
I'') to the observed radio emission of Sw 1644+57 and we 
explore its implications.  We model the radio data
presented by B12 and Z13 using a minimal set of initial assumptions and explore
different relativistic equipartition scenarios and
their consequences.  The strength of the equipartition arguments is that they
depend only on the conditions within the emitting region and they are
independent on the origin of these conditions.
As such, the results are independent of the details of the model and
serve as a useful guideline to build upon. 

In Paper I we generalize and expand the relativistic equipartition
arguments presented in Z11 (which are based on the theory presented in Kumar
\& Narayan 2009), including
several variants of the relativistic version. For each scenario, we consider 
the usual equipartition configurations, in which the energy in magnetic field 
and particles are comparable and the total energy is a minimum 
(Pacholczyk 1970; Scott \& Readhead 1977; Chevalier 1998).  This approach has
been extensively used to model the
radio observations of supernovae (see, e.g., Shklovskii 1985, Slysh 1990,
Chevalier 1998, Kulkarni et al. 1998, Li \& Chevalier 1999, Chevalier \&
Fransson 2006, Soderberg et al. 2010), it was used by Z11 in the
context of Sw 1644+57, and it was also discussed by Kumar \& Narayan (2009) in
the context of the prompt emission of GRBs (Gamma-ray Bursts).  

This paper is organized as follows. In section \ref{Equipartition} we
briefly present the results of the equipartition calculation for self-absorbed 
synchrotron relativistic sources (Paper I). In Section \ref{Swift_TDE} we
apply the theory to the radio data of 
Sw 1644+57.  We find an overall increase in energy and consider
variations on the outflow geometry with time and also deviations from equipartition
by varying the microphysics parameters to keep the energy constant
in Section 4. We discuss our results in Section \ref{Discussion} and compare
them to previous modeling. A summary and conclusions are presented in Section \ref{Conclusions}.

\section{Equipartition arguments} \label{Equipartition}

We begin with a brief summary of the main arguments presented in 
Paper I (we refer the reader to this paper for details). 
Consider a relativistic source that moves along, or close enough to,
the line of sight and that produces synchrotron emission from which we observe a peak
specific flux, $F_{\nu,p}$ at a frequency $\nu_p$.  The peak frequency is
either  $\nu_a$, the self-absorption frequency, or $\nu_m$, the
frequency at which electrons with the minimal LF radiate, that is, 
$\nu_p =$ max$(\nu_a,\nu_m)$. We assume that $\nu_p$ is smaller than the cooling
frequency, and thus we ignore the effect of electron cooling.
The system is characterized by five physical quantities: The
total number of emitting electrons within the 
observed region, $N_e$, the magnetic field in the source co-moving frame, 
$B$, the LF of the electrons that radiate at $\nu_p$, $\gamma_e$, the size 
of the emitting region, $R$, and the LF of the 
source, $\Gamma$. Since the source is moving relativistically, most of the
emission is emitted within an angle of $\sim 1/\Gamma$ with
respect to the line of sight.  Using three equations: the synchrotron frequency, the
synchrotron flux and the black-body flux, we can solve for $\gamma_e$, $N_e$
and $B$ as a function of $R$ and $\Gamma$ (see Paper I):
\begin{equation} \label{gamma_e}
\gamma_e \approx 525 \left[\frac{F_{p,mJy} \, d_{L,28}^2 \,
    \eta^{\frac{5}{3}}}{\nu_{p,10}^{2} \, (1+z)^{3}} \right] \frac{\Gamma}{f_A
  \, R_{17}^{2}},
\end{equation}
\begin{equation} \label{N_e}
N_e \approx 10^{54} \left[\frac{F_{p,mJy}^3 \, d_{L,28}^6 \,
    \eta^{\frac{10}{3}}}{\nu_{p,10}^{5} \, (1+z)^{8}} \right] \frac{1}{f_A^{2}
  \, R_{17}^{4}},
\end{equation}
\begin{equation} \label{B}
B =   (1.3 \times 10^{-2} {\rm G}) \left[\frac{\nu_{p,10}^5 \,
    (1+z)^7}{F_{p,mJy}^{2} \, d_{L,28}^{4} \, \eta^{\frac{10}{3}}} \right]
\frac{f_A^{2} \, R_{17}^4}{\Gamma^{3}},
\end{equation}
where $F_{p,mJy}=F_{\nu,p}/$mJy, the units of frequency are Hz and, throughout
the paper, we use the usual notation $Q_n=Q/10^n$. For clarity, here and 
elsewhere, the observed quantities are grouped and written between square
brackets to distinguish them clearly from
the physical parameters of the system.  The redshift
is $z$ and the luminosity distance is $d_L$. We introduced a parameter $\eta$,
which is $\eta \equiv \nu_m/\nu_a$ for $\nu_a < \nu_m$, and $\eta = 1$
for $\nu_a \ge \nu_m$. We also introduced an area filling factor,
$f_A \equiv A/(\pi R^2/\Gamma^2) \leq 1$, 
which indicates the ratio of the emitting area, $A$, to the area 
from which emission can be observed (see Paper I).

To characterize the system at the minimum total energy, we need 
two additional equations.  One equation is the total energy of
the system within the observed region.  It turns out, as in
the Newtonian case, that the electrons' energy and the magnetic field energy
are strong functions of radius, given by 
\begin{eqnarray} \label{E}
&E& = E_e + E_B \nonumber \\
&\approx&  (4.4 \times 10^{50} {\rm erg}) \left[ \frac{F_{p,mJy}^4 \,
d_{L,28}^8 \, \eta^{5}}{\nu_{p,10}^{7} \, (1+z)^{11}} \right]
  \frac{\Gamma^2}{f_A^{3} \, R_{17}^{6}} \nonumber \\
&+& (2.1 \times 10^{46} {\rm erg}) \left[ \frac{\nu_{p,10}^{10} \,
      (1+z)^{14}}{F_{p,mJy}^{4} \, d_{L,28}^{8} \, \eta^{\frac{20}{3}}}
    \right] \frac{ f_A^{4} \, f_V \, R_{17}^{11}}{\Gamma^{8}},
\end{eqnarray}
where the volume filling factor is $f_V \equiv V/(\pi R^3/\Gamma^4) \leq 1$ and
is the ratio of the volume of the emitting region, $V$, to the volume
from which emission can be observed.
The minimal total energy is found when $E_B \approx (6/11) E_e$, which yields
an estimate of the radius given by 
\begin{equation} \label{equi_radius}
R_{eq} \approx (1.7\times10^{17} {\rm cm}) \left[
  \frac{F_{p,mJy}^{\frac{8}{17}} \, 
d_{L,28}^{\frac{16}{17}} \, \eta^{\frac{35}{51}}}{\nu_{p,10} \,
    (1+z)^{\frac{25}{17}}} \right]
\frac{\Gamma^{\frac{10}{17}}}{f_A^{\frac{7}{17}} \, f_V^{\frac{1}{17}}}.
\end{equation}
The other equation relates the time since the onset of the relativistic
outflow, $t$, $R$ and $\Gamma$:
\begin{equation} \label{R-observed_time}
t = \frac{R (1-\beta) (1+z)}{\beta c},
\end{equation} 
where $\beta$ is the velocity of the outflow.  Equations (\ref{equi_radius})
and (\ref{R-observed_time}) need to be solved simultaneously to find $R$ and
$\Gamma$.  Clearly, if one finds $\beta \ll 1$, then the outflow is non-relativistic and
the solution reduces to the well-known Newtonian one. Alternatively, one can solve for $R$
assuming $\Gamma = 1$ (Newtonian case), and if the average expansion velocity of the source at
time $t$, $\bar{v} = R(1+z)/t$ is $\bar{v} \gae c$, then we are forced to
invoke the relativistic scenario.  The total energy is very sensitive to
$R$, therefore, $R_{eq}$ is a robust
estimate of $R$, unless we allow the energy to be much larger than the minimal  allowed
total energy. The absolute minimal total energy of the system within $1/\Gamma$ can be
obtained by substituting the obtained values for $R$ and $\Gamma$ back into
eq. (\ref{E}). To better understand the behavior of the total minimal energy
within $1/\Gamma$, here we present it as a function of $\Gamma$:
\begin{equation} \label{equi_energy}
E_{eq} \approx (2.5\times10^{49} {\rm erg}) \left[ \frac{F_{p,mJy}^{\frac{20}{17}} \,
d_{L,28}^{\frac{40}{17}} \, \eta^{\frac{15}{17}}}{\nu_{p,10} \,
    (1+z)^{\frac{37}{17}}} \right]
\frac{f_V^{\frac{6}{17}}}{f_A^{\frac{9}{17}} \, \Gamma^{\frac{26}{17}}}.
\end{equation}
For the case of $\nu_a > \nu_m$, there are more electrons that radiate at
$\nu_m$, which makes $N_e$ larger than in eq. (\ref{N_e}), with $\eta=1$, 
by a factor of $(\nu_a/\nu_m)^{(p-2)/2}$, where $p$ is the electron energy
distribution power-law. For this case, $E_e$ (with $\eta=1$) will be larger by this same
factor and we can self-consistently find $R_{eq}$ and $E_{eq}$ as done above. 

We also consider the possibility that the outflow contains hot protons that
carry a significant portion of the total energy.  We write the energy of
the protons as $E_p = E_e/\bar{\epsilon_e}$, therefore, the total energy in
particles is $E_e + E_p = \xi E_e$, where $\xi \equiv 1 +
{\bar{\epsilon_e}}^{-1}$. In this case, the total energy minimization 
yields $E_B \approx (6/11) \xi E_e$.  In the Newtonian case, the radius and
total minimal energy will be increased by factors of $\xi^{1/17}$ and 
$\xi^{11/17}$, whereas in the relativistic case (with 
$\Gamma \gg 1$) they are increased by factors of $\xi^{1/12}$ and 
$\xi^{7/12}$, respectively.

Finally, even though this analysis assumes equipartition, 
one can easily extend this formalism to the case when we are
far from it.  We can do so if we know the microphysical parameters, 
$\epsilon_e$ and $\epsilon_B$, the fractions of the total energy in electrons
and magnetic field, respectively.
We define a quantity $\epsilon \equiv (\epsilon_B/\epsilon_e)/(6/11)$, and 
thus the radius will be larger than $R_{eq}$ by a factor of $\epsilon^{1/17}$
and $\epsilon^{1/12}$ in the Newtonian and relativistic case (with 
$\Gamma \gg 1$), respectively. Notice that even far away from equipartition, 
the radius (and thus the LF) is a robust estimate of the true radius, since the
dependence on $\epsilon$ is extremely weak. Using eq. (\ref{E}), the total
energy will be larger than the total minimal energy by 
$E/E_{eq} \approx 0.6 \epsilon^{-0.4} + 0.4 \epsilon^{0.6}$.  Moreover, if 
$\epsilon_e + \epsilon_B < 1$, then the total energy, $E_T$, 
will be larger than in the previous equation by a factor of 
$(\epsilon_e + \epsilon_B)^{-1}$, and in this case,
roughly, $E_T/E_{eq} \approx 0.5 \epsilon_e^{-0.6} \epsilon_B^{-0.4}$.


\begin{figure*}
\begin{center}
\includegraphics[width=12cm, angle = 0]{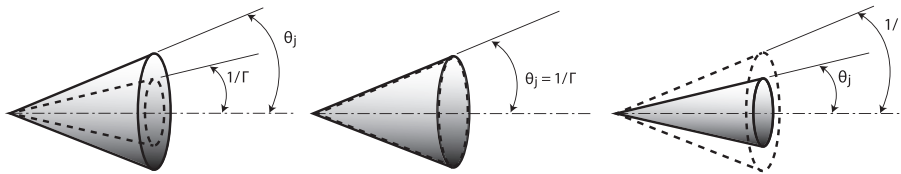}
\end{center}
\caption{Different type of relativistic outflows. From left to right:
A wide jet, where $\theta_j > 1/\Gamma$; a sideways-expanding jet, where 
$\theta_j \approx 1/\Gamma$, and a narrow jet, where $\theta_j <
1/\Gamma$. Since $\theta_j$ is unknown, for the wide jet case we
will treat the outflow as spherical and refer to isotropic quantities. 
For the narrow jet we will assume a particular value for $\theta_j$ 
and we will discuss how the results depend on the choice of $\theta_j$.
(Figure from Barniol Duran et al. 2013).}
\label{fig0} 
\end{figure*}


\subsection{The outflow geometry}

The geometry of the emitting region is an additional important factor that
controls the conditions there.
The equipartition arguments presented above can be applied to the
non-relativistic (Newtonian) case by setting $\Gamma = 1$. In the case of a 
spherical source, $f_A=1$ and $f_V \approx 1$. 
For the relativistic case, the outflow could have a wide opening angle, 
$\theta_j$, comparable or larger than  $1/\Gamma$.  In such a jet 
the observer sees only a fraction (an angle of $1/\Gamma$) of the entire jet
(this was the reason for the choice of the filling factors in the previous section).
We consider two possibilities, a ``wide'' jet, where $\theta_j > 1/\Gamma$ 
(GRB jets are believed to satisfy this condition initially)
and a ``sideways-expanding'' jet (as is the case in the late phase of a GRB 
afterglow), where $\theta_j \approx 1/\Gamma$ 
(see Fig. \ref{fig0}) .  For these two types of jets, assuming the jet is
uniform, $f_A = 1$ and $f_V=1$. For any general jet geometry or a non-uniform jet, 
$f_A < 1$ and/or $f_V < 1$. It is harder to imagine how a ``narrow'' jet with $\theta_j < 1/\Gamma$
forms as the matter will naturally tend to expand all the way to $1/\Gamma$. 
Still, in the spirit of a general equipartition approach we 
do not consider how the outflow formed and just examine what are the 
possible conditions within the emitting region. 
In the narrow jet case the observer sees the entire jet while the
jet's emission is beamed into a cone wider than the jet (see Fig. \ref{fig0}), 
thus for a uniform narrow jet $f_A = f_V = (\theta_j \Gamma)^2$.
These filling factors introduce factors of 
\begin{equation} \label{narrow_jet}
(\theta_j \Gamma)^{-\frac{16}{17}} 
\ \ \ \ \ {\rm and} \ \ \ \ \ 
(\theta_j \Gamma)^{-\frac{6}{17}},
\end{equation}
in $R_{eq}$ and $E_{eq}$, see eqs. (\ref{equi_radius}) and (\ref{equi_energy}).

It is important to note that the wide and sideways-expanding jets 
yield exactly the same results since $f_A$ and $f_V$ are identical.  
However, for a wide jet, the true number of particles and energy are 
larger than those calculated above by a factor of $4 \Gamma^2 (1- \cos\theta_j)$.  
Clearly, without additional information we cannot determine $\theta_j$, therefore,
we treat the outflow as spherical and calculate the {\it isotropic equivalent}
quantities as $N_{e,iso} = 4 \Gamma^2 N_e$ and $E_{iso} = 4 \Gamma^2
E$. Similarly, for the Newtonian spherical case, the total number of emitting particles and the
total minimal energy are a factor of 4 larger than obtained in the previous
subsection (using $\Gamma=1$), since $N_e$ and $E$ in the previous section are
obtained for a region with opening angle $\sim 1/\Gamma$.

\section{Application to the radio emission of Sw 1644+57} \label{Swift_TDE}

We apply now the formalism derived in Paper I and briefly explained in the
previous section  to the radio data of Sw 1644+57 (Z11, B12, Z13). 
This potential tidal disruption event took place at $z=0.354$ 
for which $d_L=5.7\times 10^{27}$ cm (Levan et al. 2011). We focus here on 
the radio emission that followed 
the initial soft $\gamma$-rays/hard X-ray emission. The radio data seems to be
well described by synchrotron emission with a low energy steep power-law
spectrum that requires self-absorbed emission (Z11), and a high energy
power-law spectrum for which an index of $p\approx 2.5$ can be derived
(B12, Z13).  We analyze the radio data of this
event in the context of a Newtonian spherical source and also of
the three relativistic jet types considered above: wide (isotropic), sideways-expanding
and narrow. We also separately consider the effect of including the protons' energy.
We will show that including the protons' energy
has a small effect on the physical parameters of the system and only
increases the total minimal energy. Consequently we will initially 
{\it ignore} this term for simplicity and will present its effect only 
towards the end of this section.

\begin{figure*}
\begin{center}
\includegraphics[width=12cm, angle = 0]{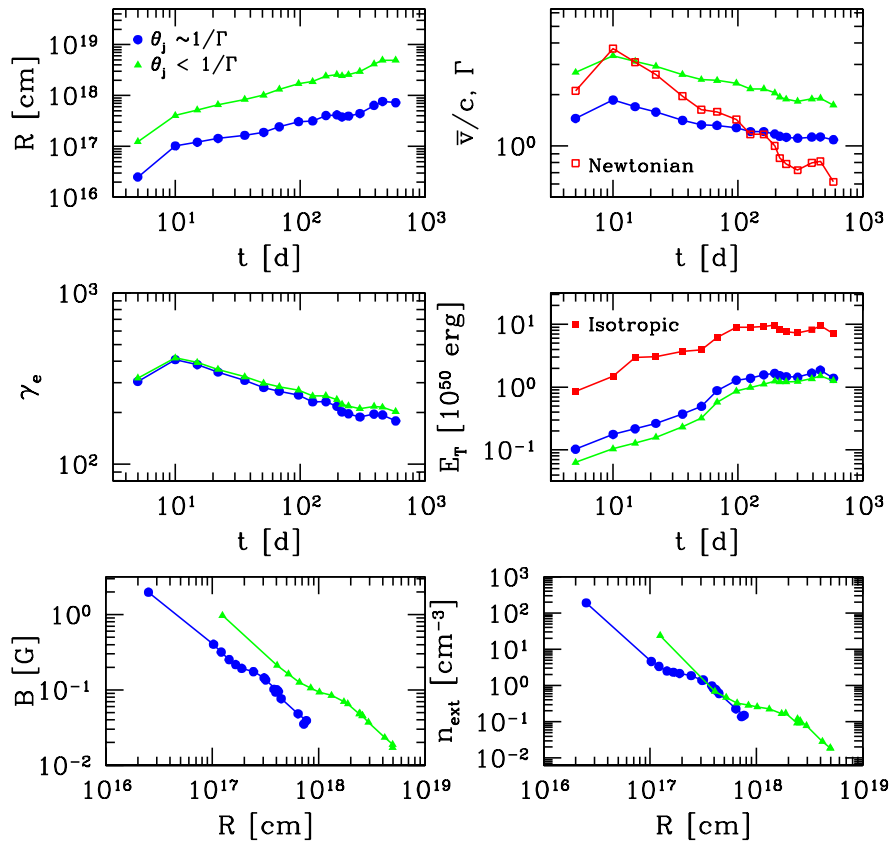}
\end{center}
\caption{Physical parameters of Sw 1644+57, {\it assuming
equipartition} (minimal total energy) 
and ignoring the protons' energy, in the following cases: a Newtonian source
(red unfilled squares), a relativistic wide jet (red squares, isotropic
quantities), a sideways-expanding jet ($\theta_j \sim 1/\Gamma$, blue circles) and
a narrow jet ($\theta_j < 1/\Gamma$, using $\theta_j = 0.1$, green triangles).  
All quantities have been calculated using the snap-shot synchrotron broad-band
spectrum fitting of B12 and Z13, that allows the determination of $\nu_a$ and hence of
$\eta$, which we call the $\eta = \eta_{obs}$ case. The magnetic field ($B$) corresponds to the
co-moving quantity.
The velocity in the non-relativistic case ($\bar{v}/c$) is the velocity
in units of the speed of light, whereas in the relativistic cases we show the
LF of the source ($\Gamma$).  No solution is found in the Newtonian spherical outflow case
as the obtained velocity exceeds the speed of light (we only
show the inferred velocity for this case in the upper right panel). 
The parameters of the wide jet and the sideways-expanding jet are identical,
except for $E_T$, for which we show the isotropic quantity.
Notice that in all cases $E_T$ increases with time by a factor of $\sim 10-20$.
The density profiles for both the sideways-expanding and narrow jets 
exhibit a flattening, but their normalizations are different (right bottom panel).}
\label{fig1} 
\end{figure*}

For each observed time, $t$, when there is an available
spectrum, we can determine $\nu_a$, $\nu_m$ and $F_p$ (see figs. 2 in B12 and
Z13). B12 and Z13 do not find a cooling frequency in their data, and infer it
to be much larger than $\nu_m$, which makes the simplification of ignoring
the electron cooling valid. For Sw 1644+57, the values of $\nu_a$ and $\nu_m$ (and hence $\eta$)
have been obtained by the snap-shot synchrotron broad-band spectrum fitting 
done by B12 and Z13 (see their tables 2\footnote{ Z13 use 
different parameters to fit the spectra for $t \ge 244$ days than B12. Z13 assume
the fraction of post-shock energy in magnetic field to be 
$\epsilon_B = 0.01$, whereas B12 use $\epsilon_B=0.1$ (they also use slightly
different values for $p$, but these do not affect the calculation).  B12 finds
that using $\epsilon_B = 0.01$ instead of $\epsilon_B=0.1$ increases both the
density and energy by a factor of $\sim 3$.  Using eqs. (1) and (2) in B12,
the $\nu_m$ ($\nu_a$) values in Z13 can be scaled to the B12 values by
multiplying (dividing) them by a factor of $\sim 1.8$ ($\sim 1.5$).}).
We will denote this case as $\eta = \eta_{obs}$ to emphasize that we are using
the observations of $\nu_a$ and $\nu_m$ to determine $\eta$. 
Using the equipartition arguments, we obtain the physical
parameters of Sw 1644+57 at each observed time in all the scenarios considered
above (see Fig. \ref{fig1}). We calculate the parameters 
of the outflow as the observed time increases from 5 to 582 days (the 
time span of the observations in B12 and Z13).  

If we assume that the radiating particles are the external medium
particles that have been swept-up by the relativistic outflow, 
and if we assume that all electrons are radiating (one
can envision a scenario in which only a fraction of them are emitting),
then we can determine the number density of the external medium, $n_{ext}$.  
The number density of radiating particles in the outflow (in the lab frame), 
$n_e = N_e/V$, is related to $n_{ext}$ by $n_e = 4 \Gamma^2 n_{ext}$ 
(Blandford \& McKee 1976), which yields 
$n_{ext} = N_e /(4 f_V \pi R^3/\Gamma^2)$. We also notice that for the
wide jet case, where we calculate isotropic quantities, 
$n_{ext}$ is the same as in the sideways-expanding case, since
$n_e$ remains the same in both cases.


The results of the equipartition calculation are
presented in Fig. \ref{fig1}.  We find no consistent Newtonian spherical equipartition
solution.  The derived radii imply an expansion
velocity larger than the speed of light (see Fig. \ref{fig1}). Therefore, we
consider only the relativistic scenarios. The relativistic sideways-expanding
jet yields $\Gamma \sim 2$, and it decreases with time.  The mildly
relativistic nature of this jet implies that the ejecta opening angle is
large, $\theta_j \sim 1$, and the outflow is almost spherical.  We also
consider a wide jet.  We show in Fig. \ref{fig1} the 
isotropic total energy in the wide jet case, which is the only
different quantity (thus the only one we plot for this case) 
between the sideways-expanding jet and the wide jet, as
explained above. Note that in the wide jet case the isotropic equivalent energy is comparable to 
the true energy, since $\theta_j \gae 1$. To account for the possibility of a narrow jet, we consider 
a relativistic jet with $\theta_j = 0.1$ (as in B12). We find 
$\Gamma \sim 3$, and it decreases with time. 
The radius in this case is larger than the sideways-expanding jet emission
radius by about an order of magnitude since 
$R \sim \theta_j^{-16/17} \Gamma^{-6/17}$, see eq. (\ref{narrow_jet}), which leads to a larger 
$\Gamma$ and to a significantly lower magnetic field and 
external density compared with the sideways-expanding jet case. We will discuss the effect of
varying $\theta_j$ in Section \ref{Varying_angle}. 

B12 find $\eta \sim 5$ and $\eta \sim 20$ at 5 and 10 days, respectively. 
It then monotonically decreases to $\eta \sim 1$ during the time span
of the observations. The sudden increase in $\eta$ by a factor of 
$\sim 4$ between 5 and 10 days (due to a sudden drop in $\nu_a$)
produces a discontinuity in all the trends we 
observe in the physical parameters in Fig. \ref{fig1} during this time span. 
We find that this is a
general feature of all jet types and it is seen in all physical parameters.
In particular, between 5 and 10 days we observe a slight increase in LF. 
This increase in LF is troublesome for a standard afterglow interpretation of 
the radio data. This sudden increase in $\eta$ between 5 and 10 d
stresses the sensitivity of the results on the interpretation of the
observations; we will discuss this more in Section \ref{Eta_1}.  
Keeping this in mind, in the following, we focus on the behavior for 
$t\gae 10$ days.

The overall trends for $10$ d $\lae t \lae 200$ d can be explained as follows.
B12 find $\nu_m \propto t^{-0.9}$, $\nu_a \propto t^{-0.2}$, which yields
$\eta \propto t^{-0.7}$; they also find that the flux is roughly constant.
We find that $\Gamma$ is slightly larger but not much larger than unity. 
Using the approximate behavior seen for $\Gamma$, $\Gamma \sim t^{-0.2}$ 
(for the sideways-expanding jet), we use (\ref{equi_radius}) to 
obtain $R \sim t^{0.3}$, and with it, we find 
$\gamma_e \sim t^{-0.2}$, $B \sim R^{-1.2}$ and $n_e \sim R^{-1.1}$, which are
the approximate trends seen in Fig. \ref{fig1}. A similar analysis can be done for the
narrow jet.  Interestingly, for {\it both} jets, the sideways-expanding (or wide) and
narrow one, the density profile shows a flattening. This flattening can be
explained by the fact that the radius varies slowly with time due to the
decrease of $\eta$, since $R \propto \eta^{\frac{35}{51}}$, see eq. 
(\ref{equi_radius}).

For all the scenarios presented above: Wide, sideways-expanding and narrow relativistic 
jets, the minimal total energy increases almost linearly with time by a factor
of $\sim 10-20$ for the first $\sim 200$ days.  After $\sim 200$ days, the total
minimal energy displays a plateau. 

\subsection{Different opening angles} \label{Varying_angle}

The narrow jet results presented above depend on the choice of the opening
angle $\theta_j$. The results for different values of $\theta_j$ can be scaled
for this value. The radius is
$R\propto \theta_j^{-16/17} \Gamma^{-6/17}$, 
and the total minimum energy is 
$E \propto \theta_j^{-6/17} \Gamma^{-32/17}$, see
eq. (\ref{narrow_jet}). A different choice of $\theta_j$ will lead to a
different value of $\Gamma$, since $R \sim t\Gamma^2$ for $\Gamma \gg 1$, see eq. 
(\ref{R-observed_time}).  With this, we find 
$E \sim \theta_j^{2/5} \sim \Gamma^{-1}$ and $R\propto \theta_j^{-4/5}$.
These allow us to determine
how the radius, LF, and total energy would vary if we choose a different 
$\theta_j$. A smaller $\theta_j$ would increase $R$, but it
would decrease $E$ and increase $\Gamma$; however, the dependence of 
$E$ and $\Gamma$ on $\theta_j$ is not very strong.  Overall varying $\theta_j$ has
the effect of shifting the curves in Fig. \ref{fig1}, but their shapes are preserved.

\subsection{The protons' energy} \label{Hot_protons}

\begin{figure*}
\begin{center}
\includegraphics[width=12cm, angle = 0]{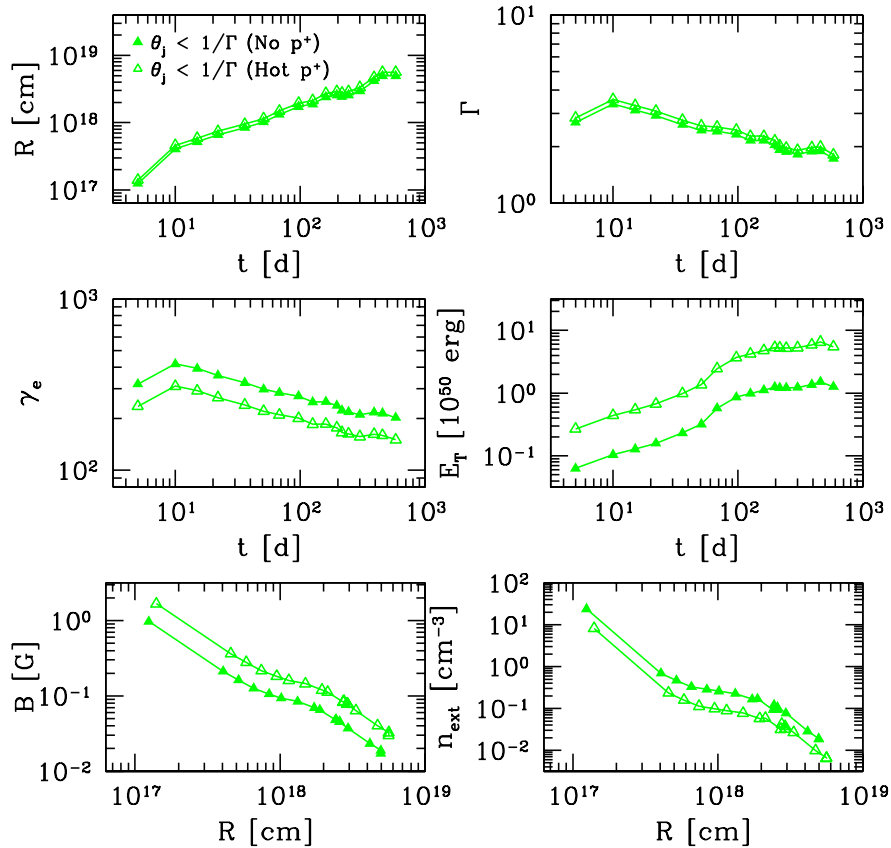}
\end{center}
\caption{A comparison of the solution with (unfilled triangles) and 
without (filled triangles) inclusion of the protons' energy for the narrow jet
case (with $\theta_j = 0.1$). The inclusion of the protons' energy increases the total
minimal energy by a factor of $\sim 5$; the rest of the parameters are shifted by only a factor of 
$\lae 3$. Similar results can be found for the case of a sideways-expanding
(or wide) jet.}
\label{fig2} 
\end{figure*}

If protons are present in the outflow, then their contribution to the total
energetics should be taken into account.  Fig. \ref{fig2} depicts, for the
narrow jet case, the effect of having 
hot protons with ten times more energy than the electrons, 
that is $\bar{\epsilon_e}=0.1$ ($\xi=11$).  
In this case the radius estimate increases only by a factor of $\sim 1.2$ 
and the total minimal energy increases by a factor of $\sim 4-5$.  All other
parameters in Fig. \ref{fig1} are only shifted by a factor of
$\lae 3$, but their behavior with time (or radius) is unchanged. This is
because of the very weak dependence of radius on $\bar{\epsilon_e}$.  


\subsection{The case of $\eta = 1$} \label{Eta_1}

\begin{figure*}
\begin{center}
\includegraphics[width=12cm, angle = 0]{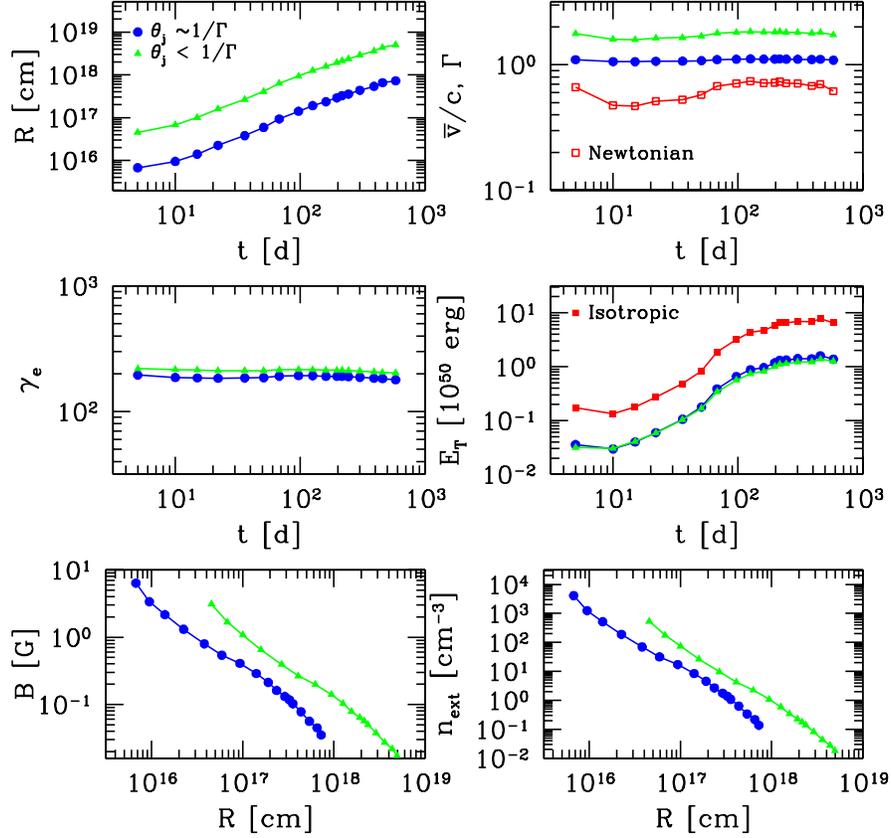}
\end{center}
\caption{Physical parameters of Sw 1644+57, {\it assuming equipartition}
and ignoring the protons' energy (same symbols as in Fig. \ref{fig1}),
using the observed peak flux and peak energy of the spectra in B12 and Z13 
and assuming $\nu_a \approx \nu_p$, that is, 
$\eta = 1$.  The resulting LF in the sideways-expanding case is close to unity, therefore, 
the physical parameters of the Newtonian case are
very similar to this case (for this reason, we chose to only show the
Newtonian velocity) and the Newtonian total minimal energy is also similar 
to the isotropic (wide jet) case. Notice that in all cases the total energy 
increases with time by a factor of $\sim 30$.  
Here, the density profile is $\propto R^{-2}$, whereas the $\eta =
\eta_{obs}$ case shows a weaker decrease with radius and exhibits a plateau.}
\label{fig3} 
\end{figure*}

To emphasize the importance of detailed radio observations that determine the
details of the spectrum, we consider the case when $\nu_a$ cannot be determined 
and set $\eta = 1$, that is, we assume that $\nu_a \approx \nu_p$ 
as done in Z11.  This is clearly an approximation that is useful when 
there is not enough data to constrain the value of $\nu_a$. 
The results for this case can be found in Fig. \ref{fig3}.  

Since $R \propto \eta^{\frac{35}{51}}$, see eq. (\ref{equi_radius}), 
assuming $\eta = 1$ results in slightly lower initial values of 
$R$ compared with the $\eta = \eta_{obs}$ case. Consequently, the LF is also 
lower (which allows for a Newtonian solution, although with very high
velocities), and both the magnetic field and the external density inferred 
are larger. Also, since $\eta_{obs}$ decreases to unity over time, 
it can be seen that the $\eta = \eta_{obs}$ solution relaxes 
to the $\eta = 1$ solution over time (see Figs. \ref{fig1} and
\ref{fig3}); this explains why $R$ varies more slowly with time
in the $\eta = \eta_{obs}$ case compared with the $\eta = 1$ case.

The main differences between the $\eta = 1$ and the $\eta = \eta_{obs}$
solutions are the following.  1. The first allows for a Newtonian solution, 
although with a high velocity, whereas the second does not. 
2. The first shows a density profile with $\propto R^{-2}$, 
whereas the second shows a weaker decrease with radius and exhibits a plateau.
3. The first shows an almost constant LF with time, whereas the second
results in a LF that decreases with time. On the other hand, the main similarity
between these two cases is that both show an increase in the
total minimal energy over time and that the energy reaches a plateau.
Using the values of $\nu_a$ in the analysis should provide a better 
estimate of the physical parameters of Sw 1644+57, therefore, we focus the 
rest of the analysis on the $\eta = \eta_{obs}$ case (otherwise noted).


\section{Alternatives to the energy increase}

In all cases discussed so far, the puzzling increase of energy over time,
which is incompatible with the X-ray emission, is a clear robust
characteristic of the equipartition solution. It arises in all solutions, and
seems unavoidable. We turn now to explore alternatives to this feature.

\subsection{The outflow geometry} \label{Geometry}

One way to avoid the increase in energy is to allow the geometry 
to vary with time, so as to counteract the energy increase.
The main idea is to let $f_A$ and/or
$f_V$ vary so that the energy in eq. (\ref{equi_energy}) remains constant.  
In the relativistic case, an observer can only see a region within 
$\sim 1/\Gamma$ from the line of sight, therefore, any change in the observed
area should occur within this region, otherwise it would not be detected.

There are numerous  ways to vary the geometry.  Here, we consider  two
as a demonstration of the behavior of the system. We can fix $f_A$ and 
let $f_V$ vary with time. This can happen if the width of the ejecta 
varies with time while we keep $\theta_j$
constant.  Alternatively we can let both
$f_A$ and $f_V$ vary with time, as will be the case if  the
outflow expands sideways by varying $\theta_j$ in the narrow jet case.  
In the following we explore how $f_V$ or $\theta_j$ should vary in these two 
cases in order to keep the total energy constant. 

The radius depends extremely weakly on $f_V$. Varying $f_V$ leaves, effectively, 
the radius unchanged (see eq. (\ref{equi_radius})), and consequently also
leaves $\Gamma$ unchanged. The minimal total energy, eq. (\ref{equi_energy}), 
depends on $f_V$ as $E\propto f_V^{6/17}$. 
This means that in order that the total energy will not  increase by the
factor of $\sim 20$ we found in the
previous section, $f_V$ needs to decrease by the very extreme 
factor of $\sim 5000$ during the time span of the observations.  

The radius in the narrow jet case behaves like  
$R\propto \theta_j^{-16/17} \Gamma^{-6/17}$, 
see (\ref{narrow_jet}). Therefore, a variation of  $\theta_j$ with time will lead to 
a large departure from the results obtained when $\theta_j$ was kept fixed. 
The total minimal energy is
$E \propto \theta_j^{-6/17} \Gamma^{-32/17}$, see
eq. (\ref{narrow_jet}). As we vary $\theta_j$ the radius changes and so will
$\Gamma$, see eq. (\ref{R-observed_time}), and, as obtained before,
$E \sim \theta_j^{2/5} \sim \Gamma^{-1}$.  Therefore, to avoid the
observed increase in total energy by a factor of $\sim 20$,  
$\theta_j$ has to decrease by $\sim 2000$ and $\Gamma$ has to increase by a
factor of 20. Clearly this scenario is unreasonable. There
is no physical reason why $\theta_j$ should decrease by such a large factor. 
Additionally, the increase in $\Gamma$ is strange, as we expect the outflow 
to decelerate rather than to accelerate at these late stages of its evolution.  

\subsection{Beyond equipartition} \label{Non-equipartition}

The total energy increases in all the equipartition scenarios presented
above. This is an inevitable result if we assume equipartition. We turn now to 
consider a scenario in which we force the total energy to be a constant and
calculate the physical parameters of the system for this to occur. We 
will again ignore the protons' energy, since it only changes the parameters 
by a small factor as explained earlier.

We use eq. (\ref{E}) together with 
eq. (\ref{R-observed_time}) to solve for $R$ and $\Gamma$ that 
keep the total energy constant.  We then calculate $\gamma_e$, $N_e$ and $B$, 
and determine the microphysical parameters 
$\epsilon_e=E_e/E$ and $\epsilon_B=E_B/E$ as a function of time 
($\epsilon_e$ is the fraction of the {\it total} energy that
goes into electrons, whereas $\bar{\epsilon_e}$ instead is the fraction of the 
protons' energy that goes into electrons). 

The total energy should be larger than the
minimal total energy obtained in Fig. \ref{fig1}.  
The minimal $E$ increases until $\sim 2\times 10^{50}$ erg, therefore, we
choose this value as the fixed energy. This means that towards the end of the 
observations we will reach roughly an equipartition solution, 
whereas earlier on there will be a significant departure from equipartition. 

\begin{figure*}
\begin{center}
\includegraphics[width=12cm, angle = 0]{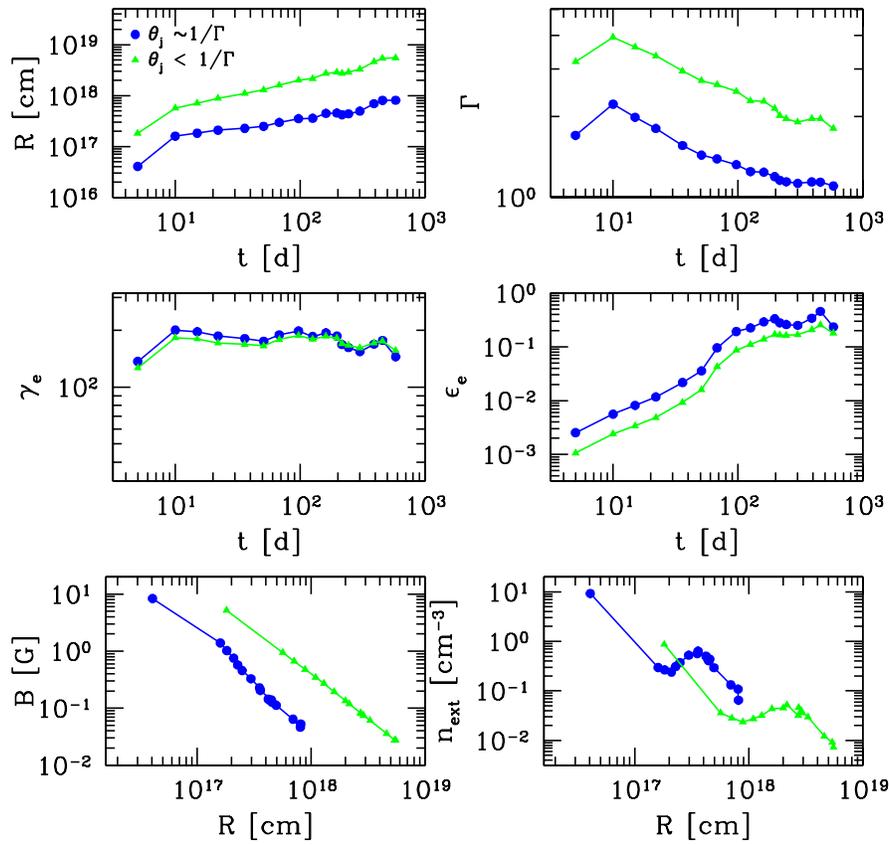}
\end{center}
\caption{Physical parameters of Sw 1644+57 for a sideways-expanding
jet (solid circles) and for a narrow jet with $\theta_j=0.1$ (solid
triangles) for a {\it fixed} total energy at $E = 2 \times 10^{50}$ erg 
(and ignoring the protons energy). All symbols and labels are the same as in
Fig. \ref{fig1}. In these cases, initially $\epsilon_B \sim 1$ and it slowly decreases with
time (not shown) as $\epsilon_e$ increases with time as in the right middle
panel (see text). Notice that the density profile exhibits a ``bump'' in both
relativistic cases (right bottom panel).}
\label{fig4} 
\end{figure*}

\begin{figure*}
\begin{center}
\includegraphics[width=12.1cm, angle = 0]{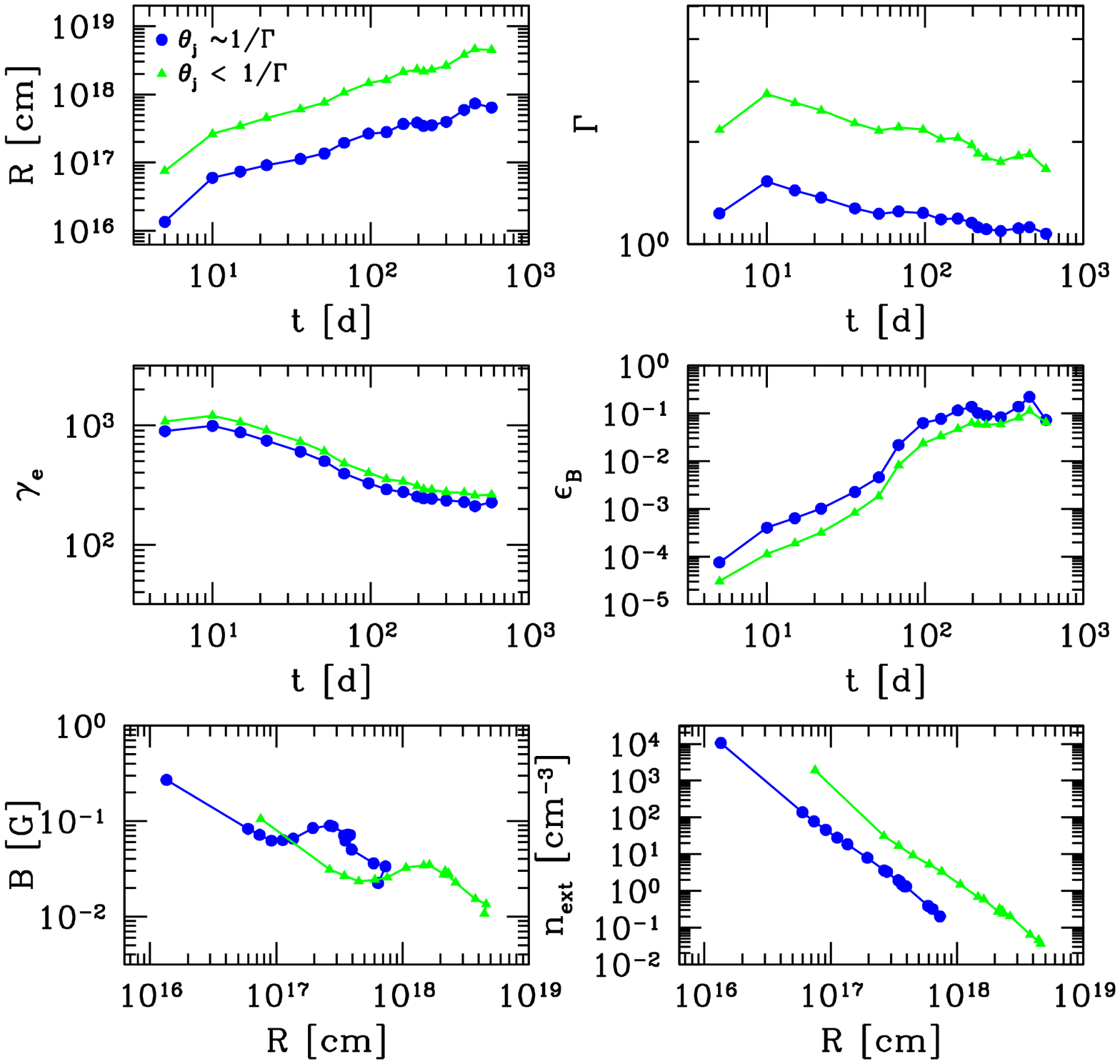}
\end{center}
\caption{Physical parameters of Sw 1644+57 for a sideways-expanding
jet (solid circles) and for a narrow jet with $\theta_j=0.1$ (solid
triangles) for a {\it fixed} total energy at $E = 2 \times 10^{50}$ erg 
(and ignoring the protons energy). All symbols and labels are the same as in
Fig. \ref{fig1}. In these cases, initially $\epsilon_e \sim 1$ and it slowly decreases with
time (not shown) as $\epsilon_B$ increases with time as in the right middle
panel (see text). Notice that the magnetic field as a function of radius
exhibits a ``bump'' in both relativistic cases (left bottom panel).}
\label{fig5} 
\end{figure*}

There are two solutions for each one of the scenarios considered:
sideways-expanding and narrow jet (as mentioned before, the wide jet case
yields same results as the sideways-expanding one). 
The first solution is a ``Poynting flux dominated'' solution, in which initially
$\epsilon_B \sim 1$ and $\epsilon_e \ll 1$, namely most of the energy 
is carried by the magnetic field at the onset of the observations.  Eventually,
$\epsilon_e$ ($\epsilon_B$) increases (decreases) with time 
(see Fig. \ref{fig4}). The second solution is a ``baryonic'' solution,   
$\epsilon_e \sim 1$ and $\epsilon_B \ll 1$ at the beginning 
and then $\epsilon_B$ ($\epsilon_e$) increases (decreases) with time (see
Fig. \ref{fig5}).


The total energy was fixed to a value that is larger than the equipartition
total minimal energy.  If the solution is of the magnetically dominated kind, that is, most
of the energy is in the magnetic field at the beginning of the observations, then
the initial radius needs to be larger, but only slightly larger, than $R_{eq}$ (see
Fig. \ref{fig4}).  This is because $E_B \propto R^{11}$.  
For this reason, $\epsilon_e$ will start out very small, and will later
increase.  A similar situation arises in the baryonic solution.
In this case, the radius will be initially slightly smaller than $R_{eq}$, since
$E_e \propto R^{-6}$.  However, the dependence of $E_B$ on radius is so
much stronger than the one of $E_e$, that the initial $\epsilon_B$ will be
even smaller than the initial value of $\epsilon_e$ in the magnetically dominated solution.

The equipartition solutions presented in Fig. \ref{fig1} showed two features: 
An increase in $E$ with time and a flattening in the density profile.  When
fixing the total energy, as done in this subsection, these two particular features translate to: 
1. An increase in $\epsilon_e$ ($\epsilon_B$) and 2. A small ``bump'' in the external density
(magnetic field); see Fig. \ref{fig4} (\ref{fig5}). These
``bumps'' are small, the external density and magnetic field increase
only by a factor of $\lae 4$, still they pose a puzzle to some models as it is not clear why
such bumps would arise. 


Choosing a total energy, $E$, larger than the value used above would affect
the results of the sideways-expanding (and wide) jet the following way.  
In the magnetically dominated (baryonic) solution, both the radius
and LF would increase (decrease) as $R \propto E^{1/7}$ and $\Gamma \propto
E^{1/14}$ ($R \propto E^{-1/5}$ and $\Gamma \propto E^{-1/10}$), for the
highly relativistic case\footnote{Recall from the previous section that no
  Newtonian solution was found for $\eta = \eta_{obs}$, because the outflow's
  velocity, $\bar{v}$, was found to exceed $c$ (see Fig. \ref{fig1}). 
  For the baryonic solution the outflow's velocity decreases with increasing
  $E$. A non-equipartition Newtonian solution {\it can} be found only if $E \gae 3 \times
  10^{54}$ erg, when $\bar{v}$ is less than a fraction of $c$.  However, we do
  not present this result here since the value of $E$ is excessive.}.  
As done in the previous section, we chose $\theta_j=0.1$ to display our
results for the case of a narrow jet in 
Figs. \ref{fig4} and \ref{fig5}. Choosing a different value of $\theta_j$ 
or a different value of total energy, $E$, would yield 
$R \propto \theta_j^{-5/6} E^{1/12}$  and $\Gamma \propto \theta_j^{-5/12} E^{1/24}$  
($R \propto \theta_j^{-3/4} E^{-1/8}$  and $\Gamma \propto \theta_j^{-3/8} E^{-1/16}$)
in the highly relativistic limit for the magnetically dominated (baryonic) case.

The same general conclusions in this subsection apply if we include 
the effect of the protons' energy.  As noted earlier, the only effect of including the
protons energy in the equipartition calculation is to increase the total
energy by a constant factor.  If we include the protons' contribution and 
increase the chosen total energy in Figs. \ref{fig4} and \ref{fig5} by this
factor, then the parameters in these figures remain unchanged.  
The precise values of $\epsilon_B$ and $\epsilon_e$ are modified, but their
general trend with time is preserved. In this case, a time-varying
$\epsilon_p$ is introduced, which is the fraction of the protons energy to the total energy. 

\section{Discussion} \label{Discussion}

We have applied general equipartition considerations (Paper I)
to the radio data of the tidal disruption event candidate, Sw 1644+57, and obtained estimates 
of the conditions within the emitting region. Before  exploring the 
implications of these finding to  different astrophysical scenarios,
we remind the reader that  the
equipartition considerations are based on several key assumptions.  
In particular, we assume that 1. The jet moves along, or close
enough to, the line of sight and 2. The electrons emit only synchrotron radiation and they 
are slow cooling. The first assumption is reasonable in all models that assume that 
the jet producing the radio outflow is aligned with the jet producing the
X-ray emission and, mainly, in all models in which the radio emission is
produced by the same jet that emits the X-rays, otherwise the X-ray energy budget
would be too large (see also below). The second assumption is natural in view
of simplicity, but see Kumar et al. (2013) for a possible significant deviation from it. 

Within the limits of these assumptions, the equipartition 
arguments yield a robust estimate of the radius of emission, and thus the 
LF, and of the minimal total energy. The actual total energy could be, of course,
much larger if the emission is not maximally efficient.  
In models with fixed values of $\epsilon_e$ and $\epsilon_B$ 
different from the equipartition values of $\epsilon_e = 0.65$ and 
$\epsilon_B = 0.35$, the radius (and LF) are only changed by a small factor
and the energy is increased by a constant amount, as explained in Section 
\ref{Equipartition}.  

The main difference between the LF of the wide 
(or sideways-expanding ) jet and the narrow jet is the fact
that the former shows a lower value of $\Gamma$. For example, at 10 days, 
$\Gamma \sim 2$ for a wide  jet, but $\Gamma \sim 3.5(\theta_j/0.1)^{-2/5}$ 
for a narrow jet with $\theta_j \lae 0.1$; note that lower values of $\theta_j$ yield
larger values of $\Gamma$ (see Section \ref{Varying_angle}). 
Thus a narrow jet, being more relativistic, might be consistent 
with being the same jet that produced the early X-ray emission (MGM12, B12). A wide jet is 
actually quasi-spherical and mildly relativistic, as suggested by Z11. 

\subsection{Interpretation of the observed radio data}

We stress that the equipartition results depend strongly on the interpretation of 
the observed radio data and, in particular, on the exact determination of
$\nu_a$, the self absorption frequency. 
Different interpretations of the radio data ($\eta=1$
versus $\eta = \eta_{obs}$) yield significantly different physical parameters 
(see Section \ref{Eta_1}). In particular, an
important difference between the $\eta=1$ and $\eta = \eta_{obs}$ cases
is that the former shows an approximately constant $\Gamma$ during the time
span of the observations, whereas the latter shows a decrease of
$\Gamma$ with time.  B12 explain the energy increase by invoking
a relativistic jet that was launched with a wide range of LFs.  $\eta = \eta_{obs}$ is essential for  this
scenario in which slower material carries more energy which is injected to the shock
at a later time after the shock slows down.  While these differences stress the importance of
detailed observations and careful analysis, they also demonstrate the
sensitivity of the results to possible observational errors, or misinterpretation of the data.
B12 and Z13 do not provide estimates of the errors in $\nu_a$ and $\nu_m$ in
their spectral fitting. If these errors were to imply $\eta$ values 
between $1$ and $\eta_{obs}$, we find that  there is an ample
room for different physical models. Nevertheless we continue the analysis
using $\eta=\eta_{obs}$ that seems to employ the best use of the current data. 

Within the same context of the interpretation of the data,  
B12 find a sudden increase in $\eta$
by a factor of $\sim 4$ between 5 and 10 days. This shows up as a discontinuity 
during this time span in the trends of all physical parameters.  
This is a general conclusion, which does not depend on the type
of jet we consider.  In particular, this effect gives rise to a slight increase
in LF between 5 and 10 days, which is puzzling for a standard afterglow
interpretation. An achromatic break is
observed in the radio light curves at $\sim 10$ days (Z11, Wiersema et
al. 2012).  MGM12 have interpreted this break as the time when the reverse 
shock has completely crossed the ejecta, which marks the transition between 
two phases: 1. the phase when the reverse shock is still crossing the ejecta 
and 2. the phase when the flow settles into the Blandford-McKee self-similar 
solution.  If this interpretation is correct, then there is no reason why the 
properties of the blast wave (for example, $\epsilon_B/\epsilon_e$) should be 
the same in these two phases as considered in our work. It is possible that 
Rayleigh-Taylor instabilities can mix the ejecta and the shocked medium during 
the first phase (see, e.g., Levinson 2009, Duffell \& MacFadyen 2013), which
could in principle vary $\epsilon_B/\epsilon_e$ for $t<10$ days and cause the 
slight increase in $\Gamma$ as observed in our results (see, also, Kumar et
al. 2013 for a discussion on the observed drop in $\nu_a$ between 5 and 10
days). In addition, the achromatic break in the radio light curves is on a
similar time scale than that of the energy injection of the jet as seen in the 
X-rays.  This might suggest that both the X-ray and radio components originate 
from the same jet, since a break in the radio light curve would be
coincidental if the X-ray and the radio components originate from different 
and unrelated outflows (see Discussion in \S \ref{X_ray} and MGM12).

\subsection{Comparison with other work}

We compare our equipartition calculation for a wide jet with Z11, who 
considered a wide jet scenario (see Fig. \ref{fig3}) using here $\eta=1$
as they did. The results are consistent within a factor of $\sim 2$ for
$t=5-22$ days (see Z11 supplementary Information, table 2). 
Z11 allow for a proton energy 10
times larger than that in the electrons alone, and so the isotropic energy in
Fig. \ref{fig3} should be multiplied by a factor of $\sim 5$ (see Section
\ref{Hot_protons}). We also find that $\Gamma$ is close to unity and that it 
remains constant during this time span. When we continue the calculations to later times
we find that additional energy is needed and the overall energy increases by a factor of 
$\sim 30$ before saturating after about 200 days.  

The equipartition results for a narrow jet with $\eta = \eta_{obs}$ and 
$\theta_j = 0.1$ (see Fig. \ref{fig1})
should be compared with those presented in B12 (and Z13, see also MGM12).  
We find that the radius is identical to the one found in 
fig. 3 in B12; we also obtain a temporal behavior of 
$R \propto t^{0.6}$ for $t \gae 10$ days.  In addition, we find $\Gamma \sim 3.5$ and
a decrease to $\Gamma \sim 2$ for $t > 10$ days. These results  approximately agree
with B12 ($\Gamma$ would correspond to their value of $\Gamma_{sh}$ in their
``afterglow-like'' model).  We obtain the same
density profile as in their fig. 6 (within a factor of $\sim 3$).  This
density profile, as discussed before, shows a shallower decay compared to the
$\eta = 1$ case, and it also shows a flattening. We also find an
increase in the total energy by a factor of $\sim 20$ as found in B12 and Z13.
Z13 find an increase in the total isotropic-equivalent kinetic energy of the
jet from $\sim 10^{53}$ erg to $2 \times 10^{54}$ erg, which translates to
$\sim 5 \times 10^{50}$ erg to $\sim 10^{52}$ erg using an opening angle of
$\theta_j = 0.1$ (Z13). The overall behavior is similar and the radius
and subsequently $\Gamma$ are comparable.  The minimal energy inferred from
the  equipartition arguments is smaller by a factor of $\sim 50$ compared with the 
one derived by Z13 (see Fig. \ref{fig2}). Making use of the Z13 values of 
$\epsilon_e = 0.1$ and $\epsilon_B = 0.01$ the 
total energy would increase by a factor of $\sim 15$ (the protons energy
results in   a factor of $\sim 10$ and the ``non-equipartition" ratio of 
$\epsilon_B / \epsilon_e$ results in an additional factor of $\sim 1.5$; 
see Section \ref{Equipartition}). 
Overall, our energy estimate is smaller than the one in Z13 by a factor of $\sim 3$.   
This comparison with  previous radio modeling demonstrates  that,
indeed,  the detailed afterglow-like modeling (MGM12, B12) reduces  to the simpler
equipartition arguments  as long as  $\epsilon_e \sim \epsilon_B$. This is valid, of course
as long as the afterglow  emission is dominated by just one component: the
forward shock in this case (MGM12). 

\subsection{The X-ray data and the two possible scenarios for the radio data} \label{X_ray}

The total X-ray isotropic equivalent energy of Sw 1644+57 is
$E_{x,iso} \approx 3 \times 10^{53}$ erg (Bloom et al. 2011, Burrows
et al. 2011). Assuming that the energy released in the X-ray band is about
one third of the bolometric energy, $\epsilon_{bol} \sim 1/3$ (Bloom et
al. 2011), and that the radiation efficiency is $\epsilon_{rad} \sim 0.3$
(although it could be smaller), the total isotropic energy in the jet needed 
to produce this emission is an order of magnitude larger 
$E_{j,iso} \approx 3 \times 10^{54} (\epsilon_{bol} \epsilon_{rad}/0.1)^{-1}$
erg. The beaming corrected X-ray jet energy is 
$E_j = (E_{j,iso}/2) \max (\theta_j^2, \Gamma_x^{-2})$. With 
$\theta_j \sim 1/\Gamma_x \sim 0.1$ (Bloom et al. 2011) we find 
$E_j \approx 2 \times 10^{52} (\epsilon_{bol} \epsilon_{rad}/0.1)^{-1}(\Gamma_x/10)^{-2}$ erg.
In the following we compare this energy with the energies involved in the radio emission in the 
different scenarios. 

The radio data of Sw 1644+57 can be explained using two possible scenarios:
one involving a narrow jet and the second involving a wide flow.
Both scenarios require that the energy of the source increases by a factor 
of 10-20 from 5 to 200 days, but they differ concerning other aspects of the
solution, which we discuss now.

For a narrow jet at 10 days we find a total minimal energy of   
$E_r(10d) \sim 10^{49} (\theta_j/0.1)^{2/5}$ erg (see
Fig. \ref{fig1}, and Sections \ref{Varying_angle} and \ref{Hot_protons}).
The energy saturates after $\sim 200$ days at 
$E_r(\gae 200d) \sim 2 \times 10^{50}$ erg.  We find that the ratio of
$E_r(\gae 200d)$ to $E_j$ is 
$E_r(\gae 200d)/E_j \sim 10^{-2} (\theta_j/0.1)^{2/5}(\epsilon_{bol}\epsilon_{rad}/0.1)(\Gamma_x/10)^{2}$ 
(this ratio is, of course, even smaller by a factor of 20 for
the estimated radio energy at 10 days). 
This radio energy is only a lower limit on the total
energy.  Deviations from equipartition, in particular lower values of 
$\epsilon_e$ and $\epsilon_B$, result in higher energies and with 
a significant inefficient radio emission process one can 
increase the energy so that the radio energy is compatible with the X-ray
energy, that is, the same narrow source produces both the X-ray and radio
emission, as explained below. 

Within the X-ray emitting region $\Gamma$ must be 
$\ge 1/\theta_j$, otherwise the radiation is beamed to a larger angle than 
$\theta_j$ and the energy suppression is not by a factor $\theta_j^2/2$, 
which is crucial to reach an overall reasonable energy budget of the 
X-ray emitting jet. So unless we have an extreme efficiency in the X-ray emission, 
$E_j \approx 2 \times 10^{52} (\epsilon_{bol}\epsilon_{rad}/0.1)^{-1}(\Gamma_x/10)^{-2}$ erg 
is a rough estimate of the energy of the fast ($\Gamma \gae 10)$ moving
material, $E_{fast} \approx E_j $. Now, this energy
should be comparable to the fast moving material energy that produces the
radio at 10 days, $E_r(10d) \approx E_{fast}$. 
As found above, for a narrow jet with $\theta_j = 0.1$, we have 
$E_r(10d)/E_j \sim E_r(10d)/E_{fast} \sim 5 \times 10^{-4}(\epsilon_{bol}\epsilon_{rad}/0.1)(\Gamma_x/10)^{2}$.
With deviations from equipartition,  $E_r(10d)$ would
increase by a constant factor (see Section \ref{Equipartition}).  Thus, we
find that $\epsilon_B \approx 10^{-9} \epsilon_e^{-1.5}
(\epsilon_{bol}\epsilon_{rad}/0.1)^{2.5} (\Gamma_x/10)^{5}$ yields comparable
energies in the narrow jet as derived from the early radio data and
from X-ray data (however, note the strong dependence on different
quantities). Therefore, it is possible that the jet that
produced the radio emission and the X-ray emission are one and the same, 
as required by afterglow-like models (see, MGM12, B12). However,
there is an intrinsic problem now. 
If this is so, then we require now 20 times more energy in 
slower moving material (to produce the late radio emission) in the model where 
the energy increase is explained by an outflow with a
velocity gradient (B12), $E_{slow} \approx E_r(200d)$, and we have a puzzling
situation in which, after beaming corrections, the overall energy of the jet
is now $E_{slow} \sim 4 \times 10^{53}$ erg straining the overall
energy budget of the event. Furthermore, 
this energetic slower moving outflow should not emit any other signal 
apart from this radio emission.  We stress that within this model, comparing 
$E_j$ with $E_r(200d)$, the energy of the radio producing matter at 
$\sim 200$ days is not relevant, since energy was injected by a slow moving 
material that could not have contributed to the X-ray emission.
Unless this paradox is somehow resolved, it seems that this energy injection model is
unlikely.

Overall we see that the narrow
jet scenario  can be based on   an ``economical'' model,  which invokes one
jet for both the X-ray and radio emission. It requires that the radio emitting regions is 
very inefficient and strongly out of equipartition as
explained above.  The increase in the total energy required by the 
radio data can be explained by energy injected by slower moving material.
However, this leads to an intrinsic inconsistency as the energy of the slower
moving material, $E_{slow}$,  (needed to produce the relatively weak late
radio emission) is $\sim 20$ times larger
than the energy of the fast moving material, $E_{fast}$, 
(that produces the strong early X-ray signal).  This strains
the energy budget of the source and makes this energy injection model
questionable. Therefore, it is worthwhile to consider alternative
models in which the energy of the magnetic field and/or of the radio
emitting electrons increases with time (see Section \ref{Non-equipartition}) 
avoiding the need of an extra energy supply to the blast wave.
 
An additional question that arises in this model is: How can the jet remain narrow 
and not spread sideways? One may argue that as long as
one continues to inject energy over the dynamical time from a very narrow
region, then the emission region will appear narrow. 
However, in this case, when the energy injection 
stops, then the emission region will start to spread after a dynamical time
and  we should observe a steepening in the radio light curve.  According to
the radio modeling, the energy injection seems to stop at $\sim 200$ d; 
however, the radio light curve does not show any sign of a steepening until the latest
observations at $\sim 600$ days (Z13). 

In the context of a wide jet, the radio emitting region has a small LF even 
at early times (see Fig. \ref{fig1}). The inferred opening angle is large 
($\theta_j \gae 1$) and the radio source is quasi-spherical.  
Clearly, within this context it is meaningless to 
consider a sideway-expanding jet. Since the outflow is almost spherical, the true energy is close
to the isotropic one.  We find that the minimal isotropic energy of the radio 
source is initially $\sim 10^{50}$ erg and it increases by a factor of $\sim 10$
until $\sim 200$ days. Thus, like in the narrow jet case, also here energy has to
be added to the radio emitting source during this period. 
The radio emission 
is produced by a quasi-spherical and mildly relativistic source, while 
the X-ray emission is produced by a relativistic and collimated jet as
required by the X-ray data (e.g., Bloom et al. 2011).
As it is clear from the different geometries required for both emissions, 
this model requires two independent outflows
with the second one (the radio source) not very energetic compared to the X-ray source
(provided that the radio emission is not very inefficient).  

The overall increase in energy in the wide radio emitting component could be
explained, again, by all energy ejected right from the beginning with a velocity gradient. 
Since here the radio and X-ray sources are different, the inconsistency
discussed for the narrow jet does not exist.
Alternatively, this wide radio source could be a wind from the super-Eddington 
accretion disk that exists in the TDE at this stage (e.g., Narayan et al. 2007). 
One would expect the wind from an accretion disk to be quasi-spherical and at most mildly
relativistic.  The total energy for the radio source increases almost
linearly with time, which would mean that the wind luminosity should be 
almost constant with time.  This is somewhat at odds with the fact that the accretion
rate, while remaining super-Eddington, decreases like $t^{-5/3}$ during this
period.  However, the wind accretion disk properties still remain to be 
well-understood.     

As mentioned before, in both models an energy injection by a factor of 
$\sim 10-20$ is essential. We can consider, however, time-dependent deviations from equipartition as 
alternative to the energy increase.  In the same context of the two
scenarios presented above, instead of invoking a special mechanism to increase
the total energy in the radio source, the total energy could have been
injected initially primarily in one form (either Poynting flux or baryonic)
and later converted from this form to the other (see Section
\ref{Non-equipartition}). This can be viewed as a variation on the scenario in which 
the excess energy is injected at the beginning but at lower velocities. 
Therefore, these two options fall into a larger category of energy
injection mechanisms, in which the total energy is injected all at once
initially, and it is later dissipated by some particular mechanism.  
Here, the total energy is injected at the beginning, but it is hidden initially in an either 
predominantly Poynting flux or predominantly baryonic outflow.
As above, the ``economical'' scenario is one in which we have the same narrow
jet producing the X-ray and radio emission. The
jet, that is launched by a supermassive black hole, is expected to be magnetically dominated.
This is also indicated by the lack of SSC emission  as noted by  Burrows et al. (2011).
In their model, the X-rays are produced by the synchrotron process and the
lack of a GeV component is a testament of the expected very weak
synchrotron-self-Compton component. Therefore, within the context of a total
constant energy (and although other scenarios are possible, see Section
\ref{Non-equipartition}), the scenario in which both the X-ray and radio data are produced by
the same narrow jet, which is initially Poynting flux dominated and 
gradually converts its energy to particle energy, seems more natural. 
Nevertheless, this scenario has also problems.  The large energy observed in
X-rays requires a large fraction of the total energy in the magnetic jet,
which is originally Poynting flux dominated, 
to be deposited (through magnetic reconnection) in particles.  However, the
radio data requires a solution where $\epsilon_e$ is initially small and the 
outflow returns to being Poynting flux dominated {\it again}, which seems contrived.

The energy increase in the wide jet scenario can also be explained
by energy injected initially primarily in one form (either Poynting flux or baryonic)
and later converted from this form to the other as explained above. This is an
alternate scenario to injecting energy initially with a velocity gradient or
an almost constant luminosity from a super-Eddington accretion disk wind. 
Since in this particular scenario we invoke two different sources (one for the X-ray
and one for the radio emission), then this allows for more freedom and avoids
the problems discussed in the last paragraph for the narrow jet.

It is interesting to consider, within this time-dependent
non-equipartition model, and abandoning the assumption of ignoring the
electron cooling, the model suggested by Kumar et al (2013). According to this model,
a single narrow jet is responsible for both the X-ray and the radio. All the
energy is injected initially, as in the non-equipartition scenario considered above. 
However, the radio emitting electrons, that arise in the forward shock, do not
emit most of their energy via synchrotron. Instead these electrons are cooled by Inverse Compton 
of the X-ray photons and thus their synchrotron radio flux is strongly suppressed. 
As the X-ray flux decreases with time, the cooling mechanism weakens.  This
gives the impression of an {\it apparent} energy increase. 

Yet another alternative to energy injection could be that the jet was not pointed directly towards us 
(e.g., Granot et al. 2002), contrary to what was assumed in this work.  
As the LF decreases and the beaming cone is able to engulf the line of sight, 
the radio emission might give the impression that energy is increasing. However,
a jet that is not pointing towards to us will increase significantly the already strained 
energy budget required to produce the X-rays.  Additionally, 
we find that $\nu_m$ in Sw 1644+57 decreases shallower than expected in the 
off-axis model (e.g., Margutti et al. 2010). Thus, overall this scenario is quite unlikely.  

\section{Conclusions} \label{Conclusions}

We have applied general relativistic equipartition considerations (Paper I) 
to the radio data of the tidal disruption event candidate, Sw
1644+57, in the most natural context of standard synchrotron emission.
We have shown that this is a powerful tool that reproduces the details
of afterglow-like models in a simpler way.  It provides a robust estimate of
the radius and thus the Lorentz Factor of the radio emitting region, and it gives a
minimal total energy required to produce the observed emission.
In this context, we have considered a relativistic jet with a wide opening
angle as $\theta_j>1/\Gamma$ and a narrow one with $\theta_j < 1/\Gamma$. 
We considered two possibilities to analyze the synchrotron radio data of Sw 
1644+57 depending on the interpretation of the observed spectra. We either
take the synchrotron peak frequency to be approximately equal to the
self-absorption frequency $\eta = \nu_p/\nu_a = 1$ (Z11) or, alternatively, 
we take the results of the snap-shot synchrotron broad-band spectrum fitting 
and determine $\eta_{obs}$ (B12, Z13).  
 
In all cases the minimal total energy of the outflow required to produce the observed
radio emission (that is, energy in the magnetic field and radio emitting
electrons) increases almost linearly with time for the first $\sim 200$ days, 
and it reaches a plateau later. The increase in energy is independent of the
details of the spectra ($\eta = 1$ or $\eta_{obs}$) or of the type of jet
considered. This is a robust result that is independent of the analysis and
that every model should be able to reproduce. We rule out the possibility that 
variations in the source geometry are responsible to the apparent increase in energy.
The only alternative to this energy increase is if equipartition is not
satisfied and all the energy is somehow deployed initially in a form that does
not produce synchrotron emission early on. 

On the other hand, the details of the variation of the LF with time or the
external density profile depend strongly on the interpretation of the radio 
spectrum.  For $\eta = 1$, the LF is a constant and the external density profile is 
$\propto R^{-2}$, whereas for  $\eta=\eta_{obs}$ the LF decreases with time 
and the external density profile is flatter and it  displays a
plateau (as found by Z11 and B12, respectively).  Again, these differences are
generic and they are independent of the type of jet (although the
normalizations of these different quantities are different).  This emphasizes 
the sensitivity of this analysis to the details of the spectra and stresses 
the importance of detailed broad-band radio observations.

Two different geometries can explain the  radio observations 
Sw 1644+57:  a wide jet and a narrow one. For the first scenario, we find that
already at 5 days, when the first radio observation is available, a wide outflow is
only mildly relativistic  and hence it is quasi-spherical, $\theta_j \gae 1$. 
This requires two different sources for the radio and the
X-rays producing outflows.  Energy considerations suggest that the source of the X-rays is 
relativistic and collimated (Bloom et al. 2011, Burrows et al. 2011), 
whereas the radio source is mildly relativistic and quasi-spherical.  These differences,
and in particular the different geometries, suggest that the X-ray and radio are produced by 
two different sources. 
Within the context of an accretion scenario for the TDE, the radio-emitting outflow 
could arise from mildly relativistic winds emerging from a  super-Eddington
disk (where the relativistic narrow outflow responsible for the X-ray would 
arise from a jet that forms in this accretion process).  

A second possible scenario involves a narrow jet that is responsible both for
the X-ray and later for  the radio emission.  For a narrow jet, the radio producing
outflow is relativistic. 
This suggests that a common origin for these two sources, 
which is essential for an  afterglow-like model (MGM12,
B12), is possible. Note that we have shown here that  simple equipartition arguments are strong
and replace the need of detailed afterglow-like modeling.  
The increase in energy in the radio can be explained 
by energy ejected during the first three days, when most of the 
X-rays were emitted, but with slower velocities (B12). This flow catches up with the 
faster (but now decelerating) part of the jet at a later time. 
This interpretation has problems in its energetics when compared
to the energy budget from X-rays. In particular it requires that the energy of the slower
moving material, needed to produce the weak late radio emission, 
is larger by a factor of $\sim 20$ than the energy of the fast moving material that produces
the very strong early X-ray emission. The overall (beaming corrected) energy needed 
is more than $\sim 4 \times 10^{53}$ erg and this strongly constraints any TDE
model. This motivated us to consider models in which the energy of the
magnetic field and/or of the radio emitting electrons increases with time,
without having a continuous injection of energy to the blast wave, as
considered below.

In the context of the two scenarios presented above, there is an alternative to 
invoking special mechanisms to increase the total energy in the radio source.
It is possible that the total energy required to produce the radio emission is 
ejected initially predominantly in one form (e.g. Poynting flux) and with time it is converted from
this form to the other (e.g baryonic). This is {\it one} possibility among
many others in which all energy is injected initially and it is accessed at later times.
Since the jet that emerges from the massive black hole  is expected to be 
magnetically dominated (e.g., Burrows et al. 2011), a scenario in which the 
radio data is produced by the same narrow jet that is initially Poynting flux
dominated and gradually converts its energy to particle energy is more
natural, but also has problems. In addition, the narrowness of the jet at late
times poses a challenge to this model and all models that require a narrow
outflow. Other scenarios within this alternative are also possible. 
Within this context, we also mention the recent suggestion by Kumar et
al. (2013) that the radio emitting electrons suffer Inverse Compton cooling by
the X-ray emission. The increase in the effective $\epsilon_e$, as the X-ray
flux decreases with time and the cooling mechanism weakens, 
is responsible for the apparent energy increase in the radio emitting region.

An accurate determination of the angular size of the radio source
(or a strong limit) would allow us to discriminate between the two
scenarios presented above. This can be done because both models predict 
different angular sizes (due to their different radii).  
Moreover, a measurement of proper motion of the
radio source would also allow us to discriminate between the two models, since
both models predict different LFs. At present, only an upper limit of 0.22 mas on the angular size 
is available at $\sim 175$ days (B12).  This limit is  larger than the predicted angular sizes for 
both wide and narrow jet models at this epoch and it does not provide a meaningful
constraint on the opening angle.  Future radio observations might  be able to 
provide stronger constraints that could allow us to determine if the jet is
narrow or not, and to distinguish between the different physical models.

\acknowledgements{
We thank Paz Beniamini, Simone Dall'Osso, Jonathan Granot, Norita Kawanaka, 
Pawan Kumar, Ehud Nakar, Ramesh Narayan, Re'em Sari and Ashley Zauderer  
for useful discussions. The research was supported by an ERC advanced grant (GRB).}


\end{document}